\documentclass[prd,preprint,showpacs,groupedaddress]{revtex4-1}
\usepackage{amsmath}
\usepackage{amsfonts}
\usepackage{amssymb}
\usepackage{geometry}
\usepackage{graphicx}
\usepackage{natbib}
\usepackage[english]{babel}
\usepackage{graphicx}
\usepackage{epstopdf}
\usepackage{subfigure}
\usepackage{caption}
\usepackage{multirow}
\usepackage{indentfirst}
\usepackage{diagbox}
\usepackage{mathrsfs}
\usepackage{booktabs}
\usepackage[colorlinks,linkcolor=blue,anchorcolor=,citecolor=blue]{hyperref}

\begin{document}
	
	\title{Shadow of Schwarzschild Black Hole in the Cold Dark Matter Halo}
	\author{Shi-Jie Ma}
	\affiliation{Lanzhou Center for Theoretical Physics, Key Laboratory of Theoretical Physics of Gansu Province, Lanzhou University, Lanzhou, Gansu 730000, China}
	
	\author{Tian-Chi Ma}
	\affiliation{Lanzhou Center for Theoretical Physics, Key Laboratory of Theoretical Physics of Gansu Province, Lanzhou University, Lanzhou, Gansu 730000, China}
	
	\author{Jian-Bo Deng}
	\affiliation{Lanzhou Center for Theoretical Physics, Key Laboratory of Theoretical Physics of Gansu Province, Lanzhou University, Lanzhou, Gansu 730000, China}
	
	\author{Xian-Ru Hu}
	\email[Email: ]{huxianru@lzu.edu.cn}
	\affiliation{Lanzhou Center for Theoretical Physics, Key Laboratory of Theoretical Physics of Gansu Province, Lanzhou University, Lanzhou, Gansu 730000, China}
	\begin{abstract}
		 The Schwarzschild black hole in the Cold Dark Matter (CDM) halo is studied, and the radiation laws of the thin accretion disk near the black hole are discussed and summarized. The orbits of light around the black hole are also calculated. Additionally, using the Novikov-Thorne model's light intensity function of the thin accretion disk, it is possible to solve for the shadow created by the thin accretion disk near the Schwarzschild black hole as well as the observed luminosity of the disk.
		
		\textbf{keywords: }the CDM halo, Schwarzschild black hole; thin accretion disk.
		
	\end{abstract}
	\maketitle
	\section{Introduction}\label{sec1}
	
	Since Einstein established general relativity, it has explained and predicted many phenomena in cosmic research, leading to new developments in cosmology~\cite{p1,p2,p3,p4}. However, in recent years, a gravitational effect has been observed that cannot be explained without the presence of additional, invisible matter. To address this issue, cosmologists proposed the existence of dark matter, which is now believed to account for approximately 85 \% of the universe~\cite{p5,p6,p7}. Dark matter can't be easily observed because they may not participate in electromagnetic interaction. So we only could investigate these properties via these gravitational effect.
	
	James Peebles first proposed the Cold Dark Matter (CDM) model in~\cite{p8}. "Cold" means that dark matter moves slowly compared to the speed of light. The predictions of the CDM paradigm are basically consistent with the observation of the large-scale structure of the universe. Since the late 1980s or 1990s, most cosmologists prefer the CDM theory, especially the $\Lambda$CDM model~\cite{p9,p10}, to describe how the universe develops.
	
	According to the model of modern physical cosmology, the dark matter halo is the basic unit of cosmic structure. The hypothesis for CDM structure formation begins with density perturbations in the Universe that grow linearly until they reach a critical density. After that, they stop expanding and collapse to form gravitationally bound dark matter halos~\cite{p11}. The metric of black holes in dark matter has been studied in recent years~\cite{p12}, this provides theoretical help for us to study the spherically symmetric black hole in the CDM halo.
	
	The photon sphere, gravitational lens, black hole shadow, and other issues have received widespread attention, but we are particularly interested in black hole shadows~\cite{p13,p14,p15,p52,p53,p54}, especially shadow produced by the thin accretion disk as the light source, because thin accretion disk is so bright. Thin accretion disk has been widely studied~\cite{p16,p17,p26,p27,p28,p29,p30,p31,p32,p33,p34,p35,p36,p37,p38,p39,p40,p41,p42,p43,p44}.The Novikov-Thorne model was firstly proposed by ID Novikov and KS Thorne in 1973 to describe the thin accretion disk around the rotating black hole~\cite{p27,p28}. Surely it doesn't hinder us to use this model to research static black hole, like Schwarzschild black hole with thin accretion disk.
	
	In order to gain a more comprehensive understanding of the appearance of black holes in the CDM halo, we will take the Schwarzschild black hole in the CDM halo as an example and conduct a comprehensive analysis of shadows, photon rings, and lensing rings. In Section~\ref{sec2}, we calculate the geodesic of the Schwarzschild black hole in the CDM halo. In Section~\ref{sec3}, we calculate the light orbit of the black hole, obtain the deflection angle of the light orbit under the influence of black hole gravity, and explore the influence of relevant parameters of the dark matter halo on the light orbit. In Section~\ref{sec4}, we calculate the relationship between the observed light intensity of the observer at infinity and the light intensity of the light source near the black hole. In Section~\ref{sec5}, we use the light intensity distribution that conforms to the actual luminous law to plot the observed intensities. Finally, in Section~\ref{sec6}, we analyze the influence of the CDM halo parameter on the radiation mode. We provide a conclusion and outlook in Section~\ref{sec7}. In this paper, to simplify the calculation, we assume $G_N=M=C=1$.
	
	\section{Geodesic equations in the CDM halo}\label{sec2}
	
	In~\cite{p12}, we get the metric of Schwarzschild black hole in the CDM halo.
	\begin{equation}
		ds^2=g_{\mu\nu}dx^{\mu}dx^{\nu}=-g\left(r\right)dt^2+\frac{dr^2}{g\left(r\right)}+r^2\left(d\theta^2+\sin^2\theta d\varphi^2\right),\label{ds}
	\end{equation}
	where $g_{\mu\nu}$ is the covariant tensor of Riemann metric and
	\begin{equation}
		g\left(r\right)=\left(1+\frac{r}{R_0}\right)^{-\frac{8\pi\rho_0 R_0^3}{r}}-\frac{2}{r}.\label{gr}
	\end{equation}
	$\rho_0$ is the density of the CDM halo collapse, $R_0$ is the feature radius. It can be seen that when $\rho_0 = 0$ or $R_0\rightarrow 0$, $g \left(r\right)$ will degenerate to $1-2/r$, so the metric will degenerate to Schwarzschild metric.
	
	Under the arbitrary metric, the Lagrangian of the particle is given by 
	\begin{equation}
	    \mathcal{L}=\frac{1}{2}g_{\mu\nu}\dot{x^{\mu}}\dot{x^{\nu}},\label{lag}
	\end{equation}
	and the geodesic equation under any metric is
	\begin{equation}
		\frac{d\dot{x^{\sigma}}}{d\lambda}+\Gamma_{\mu\nu}^{\sigma}\dot{x^{\mu}}\dot{x^{\nu}}=0,\label{geo}
	\end{equation}
	where $\lambda$ is affine parameter, $\dot{x^{\mu}}=\frac{dx^{\mu}}{d\lambda}$ are the four-velocities of the physical particle or the tangent wave vectors of the light and $\Gamma_{\mu\nu}^{\sigma}$ are Christoffel symbols, $\Gamma_{\mu\nu}^{\sigma}$ are given by
	\begin{equation}
		\Gamma_{\mu\nu}^{\sigma}=\frac{1}{2}g^{\sigma\delta}\left(\partial_{\mu}g_{\delta\nu}+\partial_{\nu}g_{\mu\delta}-\partial_{\delta}g_{\mu\nu}\right),\label{Chri}
	\end{equation}
	with $g^{\mu\nu}$ is the inverse tensor of Riemann metric.
	
	For the static spherically symmetric metric, we can always limit the motion of particles on the equatorial plane, that is $\theta = \pi/2$ and $d\theta/ds = 0$, so we can get the following three equations:
	\begin{equation}
		\dot{t}=\frac{E}{g\left(r\right)},\label{dott}
	\end{equation}
	\begin{equation}
		\dot{\varphi}=\frac{L}{r^2},\label{dotphi}
	\end{equation}
	\begin{equation}
		\dot{r}=\sqrt{E^2-g\left(r\right)\frac{L^2}{r^2}+2\mathcal{L}g\left(r\right)},\label{dotr}
	\end{equation}
	where $E$ is energy and $L$ is angular momentum, both of which are conserved quantities. Eliminate the affine parameter to get
	\begin{equation}
		\left(\frac{dr}{d\varphi}\right)^2=r^4\left(\frac{1}{b^2}-\frac{g\left(r\right)}{r^2}+\frac{2\mathcal{L}g\left(r\right)}{L^2}\right)=V_{eff}\label{veff}
	\end{equation}
	where $b = L/E$ is called the impact parameter. 
	
	\section{Null geodesic and the total change of azimuthal angle}\label{sec3}
	
	To understand the relationship between all light orbits and $b$, we substitute the null geodesic Lagrangian ($\mathcal{L}=0$) into Eq.~\eqref{veff} and obtain:
	\begin{equation}
	    \left(\frac{dr}{d\varphi}\right)^2=r^4\left(\frac{1}{b^2}-\frac{g\left(r\right)}{r^2}\right).\label{drdphi}
	\end{equation}
	For simplicity in calculation, $r$ is replaced with $u = 1 / r$, the resulting orbital equation becomes
	\begin{equation}
		\left(\frac{du}{d\varphi}\right)^2=G\left(u\right),\label{dudphi}
	\end{equation}
	where
	\begin{equation}
		G\left(u\right)=\frac{1}{b^2}-u^2 \left(\left(1+\frac{1}{uR_0}\right)^{-8\pi \rho_0 R_0^3u}-2u\right).\label{Gu}
	\end{equation}
	
		\begin{figure}[htbp]
		    \centering
		    \includegraphics[width=0.95\textwidth]{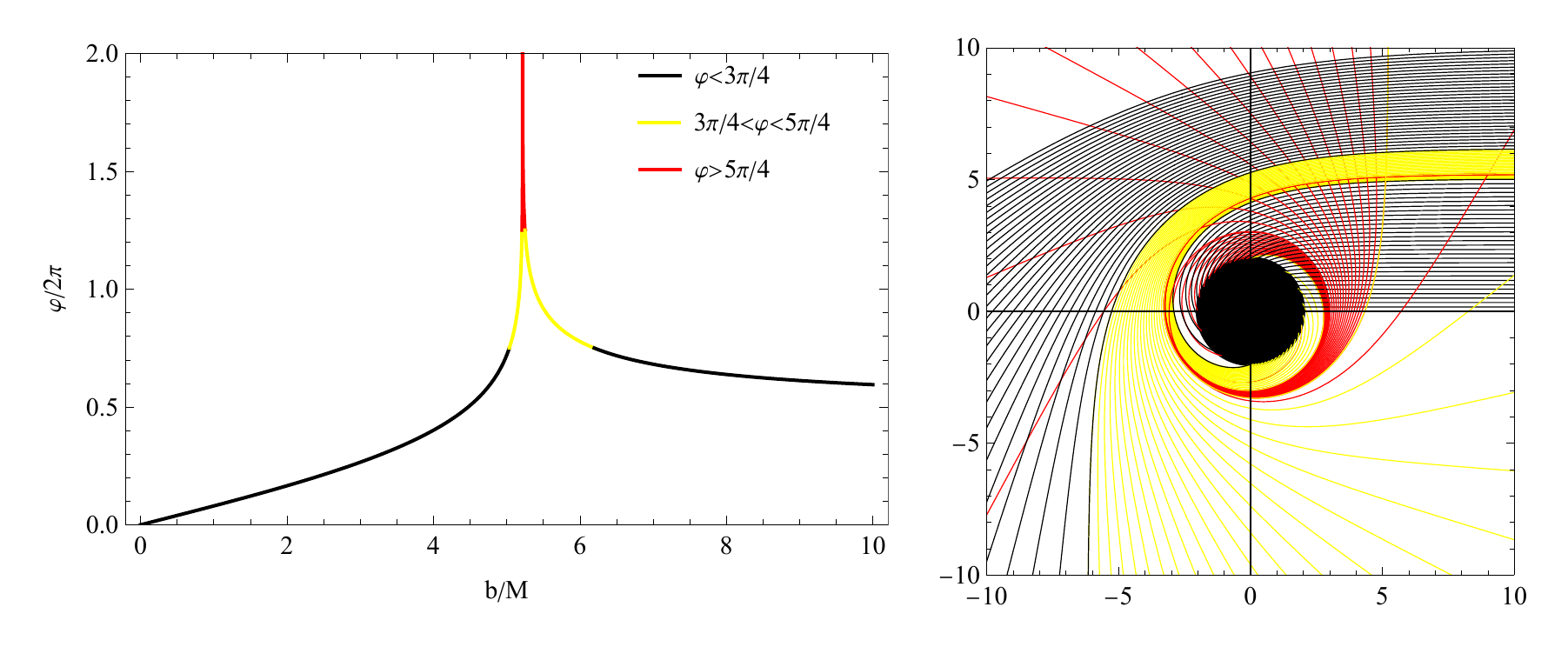}
			\caption{The relationship between $\varphi$ and $b$ (left) and the orbit of light near the black hole in the CDM halo(right). We set $\rho_0=0.1$, $R_ 0=0.1$. Different colors correspond to $\varphi < 3\pi/4$ (black), $3\pi/4 < \varphi < 5\pi/4$ (yellow) and $\varphi > 5\pi/4$ (red). On the right, we show the Euclidean polar coordinates $(r, \varphi)$. The settings of all orbit colors are the same as those in the left figure. The black hole is shown as a black disk in the middle of the image.}\label{fig1}
	\end{figure}
	
	According to Eq.~\eqref{dudphi} and Eq.~\eqref{Gu}, we can calculate the total change of azimuthal angle $\varphi$ for a certain orbit with impact parameter $b$ as follows:
	\begin{equation}
		\varphi=
		\begin{cases}
			2\int_0^{u_m}\frac{du}{\sqrt{G\left(u\right)}}& \text{ $ b>b_c $ } \\
			\int_0^{u_0}\frac{du}{\sqrt{G\left(u\right)}}& \text{ $ b<b_c $ }
		\end{cases}.\label{intphi}
	\end{equation}
	Here, $b_c$ is the impact parameter corresponding to the circular orbit of light (i.e., photon sphere), and is also the shadow radius of the black hole. We can obtain $b_c$ and the corresponding photon sphere radius $r_c$ by setting $V_{eff}=V'_{eff}=0$. At $b>b_c$, light will not fall into the black hole, and at $b<b_c$, light will fall into the black hole. $u_m$ is the turning point at $b>b_c$, corresponding to the minimum positive real root of $G\left(u\right)=0$. $u_0=1/r_0$ corresponds to the radius of the event horizon.
	
    We define $b_n^-<b_c$ and $b_n^+>b_c$, where $n$ is the number of times that light from infinity passes through the equatorial plane of the black hole. We use $b_n^\pm$ to represent the solutions of the following equation.
	\begin{equation}
		\varphi\left(b\right)=\left(n-\frac{1}{2}\right)\pi,\quad n=1,2,3\cdots\label{phib}
	\end{equation}
	According to the number of times light passes through the equatorial plane, it can be divided into direct orbit ($n=1$), lended orbit ($n=2$) and photo ring orbit ($n\geq3$).
	
	\begin{table}
		
		\begin{center}
			\caption{Various important physical parameters when $\rho_0 = 0.1$ and $R_0$  takes different values.}\label{tab1}
			\setlength{\tabcolsep}{3mm}{
				\begin{tabular}{ccccccccc}
					\hline
					$R_0$&$r_0$&$r_c$&$b_1^-$&$b_2^-$&$b_3^-$&$b_c$&$b_3^+$&$b_2^+$\\
					\hline
					0&2&3&2.8477&5.0151&5.1878&5.1962&5.2279&6.1678\\
					0.1&2.0076&3.0117&2.8588&5.0363&5.2102&5.2156&5.2506&6.1965\\
					0.2&2.0481&3.0740&2.9181&5.1515&5.3321&5.3409&5.3746&6.3575\\
					0.3&2.1375&3.2125&3.0498&5.4125&5.6096&5.6197&5.6574&6.7322\\
					\hline
			\end{tabular}}
		\end{center}

		\begin{center}
			\caption{Various important physical parameters when $ R_0= 0.1$ and $\rho_0$  takes different values.}\label{tab2}
			\setlength{\tabcolsep}{3mm}{
				\begin{tabular}{ccccccccc}
					\hline
					$\rho_0$&$r_0$&$r_c$&$b_1^-$&$b_2^-$&$b_3^-$&$b_c$&$b_3^+$&$b_2^+$\\
					\hline
					0&2&3&2.8477&5.0151&5.1878&5.1962&5.2279&6.1678\\
					0.1&2.0076&3.0117&2.8588&5.0363&5.2102&5.2156&5.2506&6.1965\\
					0.2&2.0153&3.0234&2.8700&5.0575&5.2325&5.2409&5.2733&6.2254\\
					0.3&2.0229&3.0351&2.8811&5.0787&5.2548&5.2634&5.2959&6.2543\\
					\hline
			\end{tabular}}
		\end{center}
	\end{table}
	The physical picture of this classification is clear from the orbit plots in Fig.~\ref{fig1}. Assuming that the light is emitted from the north pole (the rightmost side of the plot). When $b\rightarrow \infty$, the azimuth of light around the black hole changes to $\varphi\rightarrow\pi / 2$, so there is no corresponding $b_1 ^ +$.

	We tried to calculate $r_0$, $r_c$, $b_c$, $b_1^-$, $b_2^\pm$ and $b_3^\pm$ for different $\rho_0$ and $R_0$, and wrote them in Table~\ref{tab1} and Table~\ref{tab2}.
	
	Table~\ref{tab1} displays various important physical parameters for different values of $R_0$ when $\rho_0=0.1$. Table~\ref{tab2} displays various important physical parameters for different values of $\rho_0$ when $R_0=0.1$.
	
	Table~\ref{tab1} presents various important physical parameters for different values of $R_0$, when $\rho_0=0.1$. From the table, it is evident that all the parameters increase with an increase in $R_0$, while $\rho_0$ remains constant. Similarly, Table~\ref{tab2} shows various important physical parameters for different values of $\rho_0$, when $R_0=0.1$. The table reveals that all the parameters increase with an increase in $\rho_0$, while $R_0$ remains constant. These results suggest that light observed from different regions will be farther away from the center of the black hole as $\rho_0$ and $R_0$ increase. Additionally, Table~\ref{tab1} and Table~\ref{tab2} reveal that when $\rho_0=0$ and $R_0\rightarrow0$, the black hole degenerates to a Schwarzschild black hole in a vacuum and its parameters reach a minimum.
	
	\section{Observed specific intensities and transfer functions}\label{sec4}
	
	In general, the term 'shadow' describes how a black hole appears when illuminated from all directions. In sections~\ref{sec2} and~\ref{sec3}, we have explored the optical orbits near the black hole in the CDM halo. Now, we are ready to consider a thin accretion disk on the equatorial plane around the black hole in the CDM halo. For an observer at infinity, what is observed is more important than the distribution of light intensity.
	
	Now a simple example of a thin accretion disk around a black hole is considered, in which the radiation intensity depends only on the radial coordinates and absorption of light is neglected. We assume that the disk is isotropic within the stationary frame of the object on the stationary world line. This example serves as a model to illustrate the gravitational lensing and gravitational redshift effects. The disk is placed on the equatorial plane and we assume that the observer is located at the North Pole. $I_{\nu_{e}}^{em} \left(r\right)$ is used to express the specific intensity of emission. $\nu_{e}$ is the transmission frequency in the stationary coordinate system. The relative intensity of the light that an observer at infinity will receive is $I_{\nu_{o}}^{obs}$, $\nu_{o}$ is the frequency observed by the observer at infinity of the stationary reference system, and the frequency redshift is caused by the gravitational effect $\nu_{o}=\sqrt{g\left(r\right)}\nu_{e}$ ~\cite{p23}. Considering that there is a conserved quantity $I_\nu/\nu^3$ for a light~\cite{p24}, one can get
	\begin{equation}
		\frac{I_{\nu_{o}}^{obs}}{\nu_{o}^3}=\frac{I_{\nu_{e}}^{em}}{\nu_{e}^3},\label{eq15}
	\end{equation}
	so the relative intensity we observed is
	\begin{equation}
		I_{\nu_{o}}^{obs}=\left(\frac{\nu_{o}}{\nu_{e}}\right)^3I_{\nu_{e}}^{em} \left(r\right)=g^{3/2}\left(r\right)I_{\nu_{e}}^{em} \left(r\right).\label{Iobsnuo}
	\end{equation}
	The total intensity observed is an integral of all frequencies
	\begin{equation}
		I^{obs}=\int I_{\nu_{o}}^{obs}d{\nu_{o}}=\int g^2\left(r\right) I_{\nu_{e}}^{em}\left(r\right)d{\nu_{e}}=g^2\left(r\right)I^{em}\left(r\right),\label{Iobs}
	\end{equation}
	where $I^{em}\left(r\right)=\int I_{\nu_{e}}^{em}\left(r\right)d\nu$ is the total emission intensity at the accretion disk at $r$. 
	
	Due to the high intensity of light emitted from the accretion disk, other light sources in the environment can be ignored. If a beam of light from the observer intersects with the emission disk, it means that the intersection point will contribute to the observed brightness as a light source. As mentioned in Section~\ref{sec3}, a light ray may pass through the equatorial plane of the black hole multiple times, and each intersection with the disk will become a light source contributing to the observed intensity for that orbit. Therefore, the observed intensity is the sum of the intensities of each intersection.
	\begin{equation}
		I^{obs}\left(b\right)=\sum_n g^2\left(r_n\left(b\right)\right)I^{em},\label{Iobsb}
	\end{equation}
	where $r_n\left(b\right)$ is the so-called transfer function, which represents the radial position of the $n$-th intersection and the transmitting disk, that is, the radial coordinates of the luminous point. It should be emphasized that our model does not take into account the absorption and reflection of light by the accretion disk or the loss of light intensity in the environment, and is only an idealized representation.

\begin{figure}
		\centering
		\includegraphics{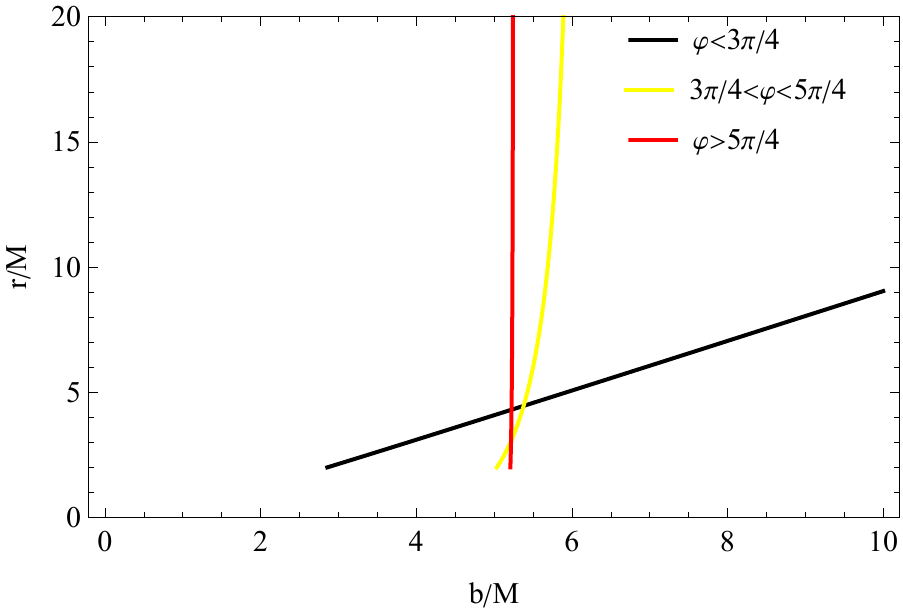}
		\caption{The first three transfer functions of black holes in the CDM halo with $\rho_0=0.1$, $R_ 0=0.1$.The three curves in the figure represent the radial coordinates of the first (black), second (yellow) and third (red) intersections with the accretion disk.}
		\label{pr-b}
	\end{figure}

	We denote the solution of orbit Eq.~\eqref{Iobsnuo} by $u\left(b, \varphi\right)$, so we can get the transfer functions:
	\begin{equation}
		r_n\left(b\right)=\frac{1}{u\left(b, \frac{\left(2n-1\right)\pi}{2}\right)}\quad b\in\left(b_n^-,b_n^+\right),\label{rmb}
	\end{equation}
	here we take positive infinity as  $b_1^+$. In Fig.~\ref{pr-b}, we show the first three transfer functions of the black hole in the CDM halo.
	\section{Radiation from stable circular orbit}\label{sec5}
	
	The model of thin accretion disk used in this paper meets the Novikov-Thorne model. According to assumptions of Novikov-Thorne model, the model conforms to conservation laws of energy ($\nabla_{\mu}T^{t\mu}$), angular momentum ($\nabla_{\mu}T^{\varphi\mu}$),rest mass ($\nabla_{\mu}\left(\rho u^{\mu}\right)$) of particles in an accretion disk. By using these conservation laws, one can obtain the radiant energy flux as~\cite{p27,p28}
	\begin{equation}
	    F\left(r\right)=-\frac{\mathcal{M}\Omega^{'}}{4\pi\sqrt{-g}\left(E-\Omega L\right)^2}\int_{r_{ISCO}}^{r}\left(E-\Omega L\right)L^{'}dr,\label{Fr}
	\end{equation}
	where $\mathcal{M}$ is the mass accretion rate, $g$ is the determinant of metric tensor, $r_{ISCO}$ is the radius of the innermost stable circular orbit (ISCO) of physical particles around a black hole ($2\mathcal{L}=-1$), The ISCO should satisfy $V_{eff}=V'_{eff}=V''_{eff}=0$. There is a relation $A^{'}=\frac{dA}{dr}$ for any physical quantity $A$. $\Omega=d\varphi/dt$ is angular velocity of particles in any stable circular orbits ($\dot{r}=\ddot{r}=0$), for any stable circular orbit, there are
    \begin{equation}\label{OEL}
        \begin{aligned}
            &\Omega=\sqrt{-\frac{g_{tt}^{'}}{g_{\varphi\varphi}^{'}}}\\
            &E=-\frac{g_{tt}}{\sqrt{-g_{tt}-g_{\varphi\varphi}\Omega^2}}\\
            &L=\frac{g_{\varphi\varphi}\Omega}{\sqrt{-g_{tt}-g_{\varphi\varphi}\Omega^2}}
        \end{aligned}
    \end{equation}
	
	\begin{figure}[htbp]
	    \centering
	    \includegraphics[width=\textwidth]{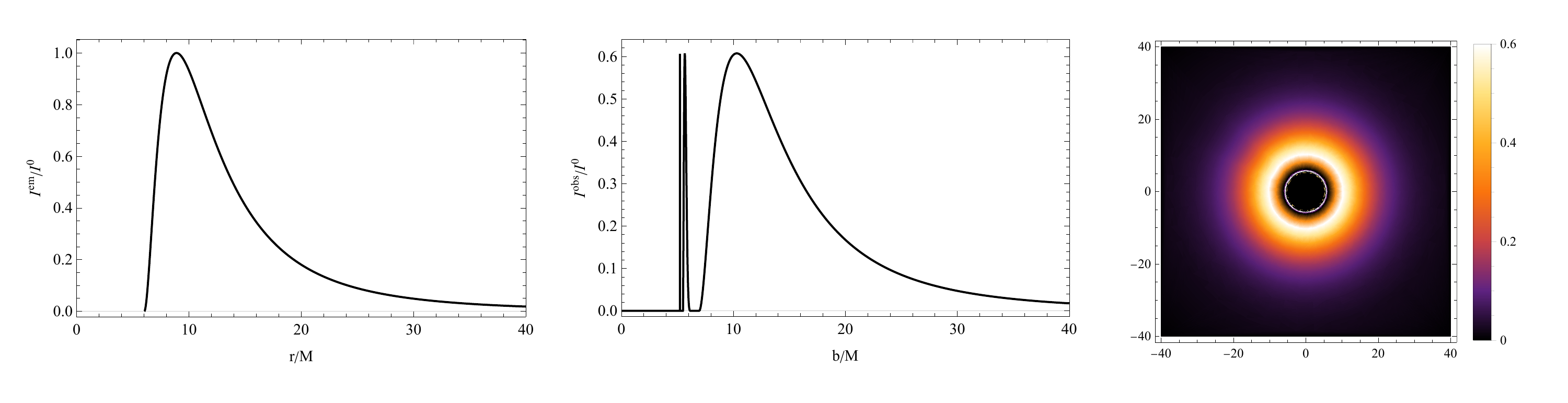}
	    \caption{Emission specific intensity diagram (left), observation specific intensity diagram (middle) and halo diagram (right) of model given by Eq.~\eqref{Fr} with $R_{0}=0.1$ and $\rho_{0}=0.1$.}
	    \label{pi}
	\end{figure}
	It is evident from Fig.~\ref{pi} that the light intensity distribution observed by an observer at infinity is composed of several intensity peaks with the same shape but different widths, and most of the observed light intensity is contributed by $r=r_{1}\left(b\right)$. Although only $n=3$ is calculated, the width of the corresponding peak narrows with the increase of $n$. As $n$ becomes larger, the intensity peaks become less significant and can be ignored.
	
    \section{Radiation flux and observed luminosity}\label{sec6}
    
	To understand the properties of the thin accretion disk near the Schwarzschild black hole in the CDM halo, the radiation flux and observed luminosity of the above radiation under different parameters have been calculated and analyzed.
	
\begin{figure}
	    \centering
	    \includegraphics[width=0.9\textwidth]{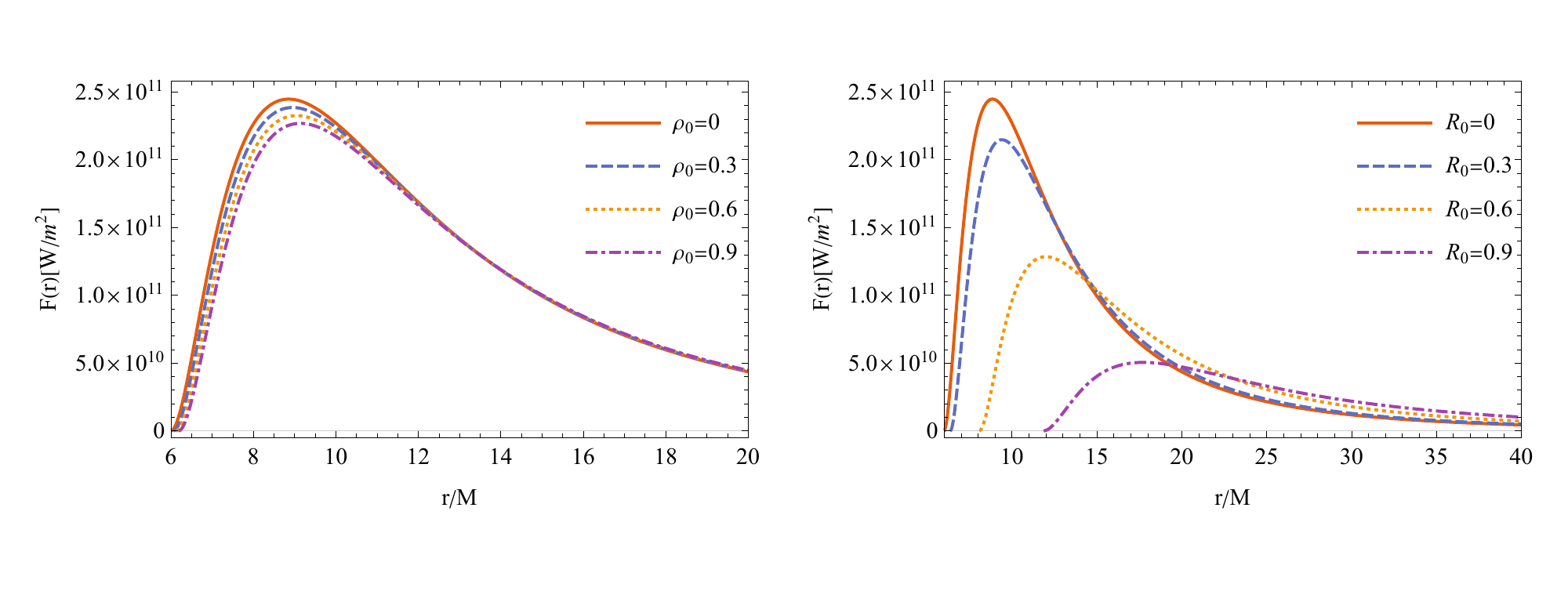}
	    \caption{The energy flux $F\left(r\right)$ of an accretion disk around the black hole in CDM halo with mass accretion rate $\mathcal{M}=2*10^{17}\, kg/s\approx 3.2*10^{-6}\, m_{sun}/year$, for different values of the parameter $\rho_{0}$ with $R_{0}=0.1$ (left) and for different values of $R_{0}$ with $\rho_{0}=0.1$(right).}
	    \label{pf}
	\end{figure}

	In addition, we have also calculated the energy flux while keeping the mass accretion rate constant and varying the other parameters, and the results are presented in Fig.~\ref{pf}. As $R_{0}$ or $\rho_{0}$ increases, the peak value and total energy flux decrease, and the radii of the corresponding peaks increase.
	
	Assuming that the accretion disk in this model is in local thermal equilibrium, we can assume that the radiation from the accretion disk is blackbody radiation. The observed luminosity of thin disks is given by~\cite{p50}
	\begin{equation}\label{lum}
	    \tilde{L}\left(\nu_{o}\right)=8\pi h\cos{\gamma}\int_{r_{in}}^{r_{out}}\int_{0}^{2\pi}\frac{\nu_{e}^{3}r dr d\varphi}{e^\frac{h\nu_{e}}{k_{B}T}-1},
	\end{equation}
	where $h$ is the Planck constant, $k_{B}$ is the Boltzmann constant, $\gamma$ is the disk inclination angle which we will set to zero. We will set $r_{in}=r_{ISCO}$, and $r_{out}$ is the outer edge of the disk, it takes as $50\,r_{0}$~\cite{p50}. $T$ is the temperature of the disk, one can get by the Stefan-Boltzmann law 
	\begin{equation}\label{sbl}
	    F\left(r\right)=\sigma T^{4}\left(r\right),
	\end{equation}
	where $\sigma$ is the Stefan-Boltzmann constant.
	\begin{figure}
	    \centering
	    \includegraphics[width=0.9\textwidth]{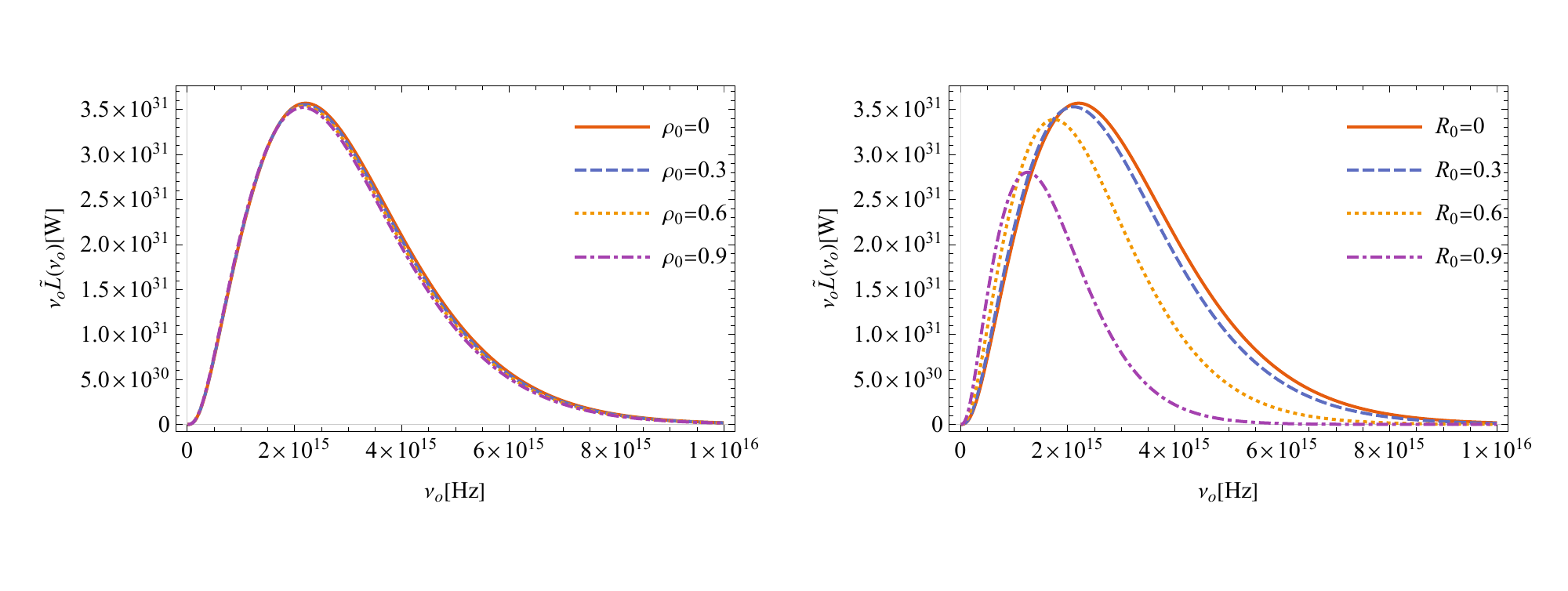}
	    \caption{The emission spectrum $\nu_{o}\tilde{L}\left(\nu_{o}\right)$ of an accretion disk around the black hole in CDM halo with mass accretion rate $\mathcal{M}=2*10^{17}kg/s$, for different values of the parameter $\rho_{0}$ with $R_{0}=0.1$ (left) and for different values of $R_{0}$ with $\rho_{0}=0.1$(right).}
	    \label{pnl}
	\end{figure}
	
	As shown in Fig.~\ref{pnl}, with the increase of $\rho_{0}$ or $R_{0}$, the accretion disk will become dimmer, which is consistent with the previous analysis, and the cut-off frequency of the maximum luminosity moves to a lower value.
	
	\section{conclusion and outlook}\label{sec7}

    First, we calculate the geodesic equation of the Schwarzschild black hole in the CDM halo and the deflection angle of the photon orbit. We also analyze the relationship between the halo radius and the CDM halo parameters. Next, we calculate the relationship between the light intensity emitted by the thin accretion disk and the observed light intensity of the observer, as well as the transfer function. The metric used in this paper conforms to the Novikov-Thorne model, and we derive the radiant energy flux according to the model. We plot the observed intensities of the radiation in the CDM halo and calculate the variation of the radiation flux and observed luminosity with the CDM halo parameters.
    
    Looking to the future, there are two potential directions for further research. Firstly, we could investigate the radiation of accretion disks in other dark matter models and black hole models to gain a more comprehensive understanding of the behavior of thin disks in different astrophysical contexts. Secondly, recent research~\cite{p25} has discovered a new shadow in a symmetrical thin-shell wormhole connecting two different vacuum Schwarzschild spacetimes. Similarly, it is possible that a new shadow may exist in a symmetrical thin-shell wormhole connecting two different Schwarzschild spacetimes in the case of the CDM halo. Therefore, it would be worthwhile to investigate the possibility of such a new shadow when considering the influence of the CDM halo.

	\section*{Conflicts of Interest}\label{sec8}
	
	The authors declare that there are no conflicts of interest regarding the publication of this paper.
	\section*{Acknowledgments}
    We would like to thank the National Natural Science Foundation of China (Grant No.11571342) for supporting us on this work.
	
	\section*{References}
	
	\bibliographystyle{unsrt}
	\bibliography{paper}
\end{document}